\documentclass{article}
\usepackage{amsmath,epsfig}

\usepackage{amsfonts, latexsym}
\usepackage{bbm}
\usepackage[latin1]{inputenc}
\usepackage[english]{babel}

\usepackage{graphicx}
\usepackage{url}
\usepackage{subfigure}
\usepackage[only,llbracket, rrbracket]{stmaryrd}
\usepackage{psfrag}

\newtheorem{prop}{Proposition}

\newtheorem{assumption}{Assumption}

\newcommand{\demo}{{\em Proof.}\,}
\newcommand{\findemo}{\hfill$\Box$}

\newcommand{\highlight}[1]{\textbf{#1}}

\newtheorem{definition}{Definition}

\newtheorem{remark}{Remark}

\title{RFID Key Establishment Against Active Adversaries}

\author{J. Bringer, H. Chabanne, G. Cohen, B. Kindarji\thanks{This work has been partially funded by the ANR T2TIT project.}\thanks{J.B. H.C. and B.K. are with Sagem Sécurité.}\thanks{H.C. G.C. and B.K. are with Telecom ParisTech.}
\\
\\ \small Sagem Sécurité, Osny, France
\\ \small Institut Telecom, Telecom ParisTech, Paris, France
}
\begin{document}

\maketitle

\begin{abstract}
We present a method to strengthen a very low cost solution for key agreement with a RFID device. 
Starting from a work which exploits the inherent noise on the communication link to
establish a key by public discussion, we show how to protect this agreement against active adversaries. For that purpose, we unravel integrity $(I)$-codes suggested by Cagalj et al. 
No preliminary key distribution is required.\end{abstract}

\section{Introduction}

Wireless communication is the source of many opportunities and challenges, one of which being its confidentiality. A convenient way to achieve confidentiality is to use cryptography, which requires for the communicating entities to detain a cryptographic key beforehand.

We focus on particular wireless devices, called RFID (for \emph{Radio-Frequency IDentification} tags). These are electronic tags made of an integrated circuit equipped with an antenna. The amount of computation possible in RFID tags is somewhat limited, due to constraints on  cost,  size and  power consumption of such devices. For that reason, protocols involving RFID devices must focus on the complexity of computation on the device side; which puts aside asymmetric cryptography.

Under this constraint, even symmetric cryptography setting must be thought thoroughly.
A solution is presented in the context of RFID in \cite{ChaFum06}, where the authors use a public discussion over a noisy channel for two wireless devices to agree on a key, and show how to realize such a protocol with low-cost tags. An eavesdropper listening to such a protocol would not gain information on the key.
As a natural extension of their work, we show how to shield a protocol of this sort in order to thwart active adversaries. The additional tools required for this additional protection are reduced to a minimal complexity.

In order to formally introduce the essential notions refered to in the rest of the paper, Section \ref{sec:formal} describes the channels that we use. Section \ref{sec:keyAgr} explains how Key Agreement through Public Discussion works. Section \ref{sec:iCodes} details $(I)$-codes, a tool that enables us to protect the Key Agreement against active adversaries. Finally, Section \ref{sec:ka-pres} presents our protocol for Key Agreement through presence. Section \ref{sec:concl} concludes.

\section{A Description of the Devices, the Channel, and the Problematic}  \label{sec:formal}

As it is often the case in cryptographic protocols, two entities Alice (\textbf{A}) and Bob (\textbf{B}) wish to communicate securely over some channel, while an adversary Eve (\textbf{E}) wants to counter their objectives, by either preventing the establishment of a key, or by discovering the key so that the communication is no longer confidential.

\begin{figure}[htb]
  \centering
	\includegraphics[width=0.48\columnwidth ]{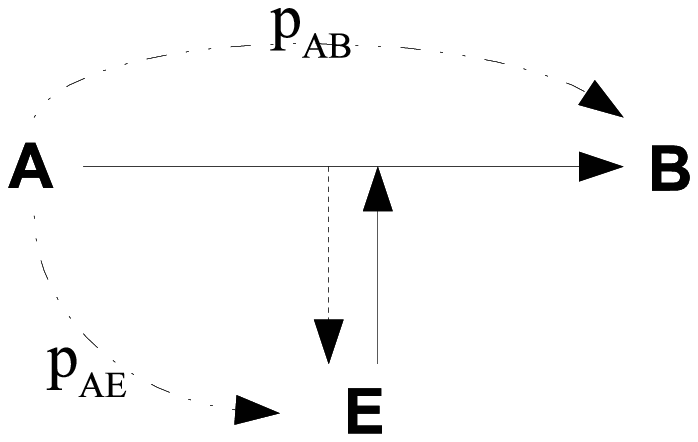} 
	\includegraphics[width=0.48 \columnwidth ]{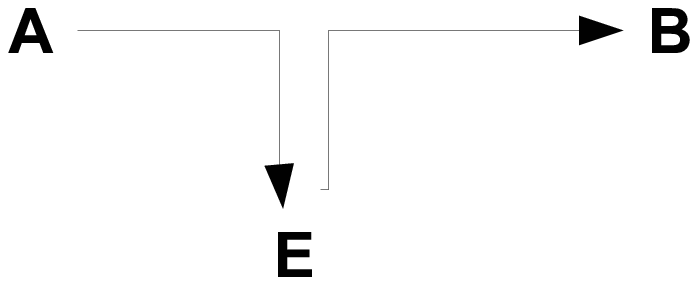}
\caption{The noisy channels $C_{p_{AB}}$, $C_{p_{AE}}$, and the noiseless channel $C_0$. }
\label{fig:noisyChan}
\end{figure}

We focus on wireless devices. This means that they communicate using radio frequency; a direct consequence is that all messages sent by these devices are public. Moreover, there is noise over the channel. This noise can be caused by
\begin{enumerate}
\item physical causes such as interferences, Doppler effect, \textit{etc.}
\item the emission of other wireless devices, that can be genuinely communicating over the same frequency, or can willingly emit in order to alter the communication.
\end{enumerate}
The presence of noise over the channel leads us to the use of Error Correcting Codes (ECC) (that enable to reduce the noise). In other terms, we have two formal channels over which the devices are able to communicate.
\begin{enumerate}
\item A noisy channel $C_p$ that inherently induces errors in the transmitted messages. We here suppose that $p_{AB}$ is a non-null error probability describing a Binary Symmetric Channel (BSC) between \textbf{A} and \textbf{B}. Moreover, we also suppose that the transmission from \textbf{A} to \textbf{E} is done through a BSC of parameter $p_{AE}$ which can be different than $p_{AB}$. (see Fig.~\ref{fig:noisyChan}).
\item A noiseless channel $C_0$ obtained by correcting errors over $C_p$.
\end{enumerate}

Both channels are public, \textit{i.e.} \textbf{E} can listen to the channel, send some messages, and even alter sent messages by adding noise.

The goal of this paper is to establish a key between \textbf{A} and \textbf{B} that is unknown by \textbf{E}. Our constraints are for \textbf{A} and \textbf{B} to be low-cost devices, which means that no sophisticated computation is allowed, and that we aim at very few logical gates to implement the protocol. As we prove in Section \ref{sec:ka-pres}, we do this by constructing a noiseless channel that detects intrusion of an active adversary, in other words, a ``shielded'' noiseless channel.

\section{Previous Results on Key Agreement} \label{sec:keyAgr}

The classical approach to key agreement by public discussion over a noisy channel was explored by \cite{ChaFum06} to apply it on low-cost devices such as RFID. Their approach follows the three steps of \textbf{Advantage Distillation} \cite{GanMau94}, \textbf{Information Reconciliation} \cite{BraSal94} and \textbf{Privacy Amplification} \cite{CacMau95}. We recall in a few lines the main ideas behind these steps.

\subsection{Advantage Distillation}

\textbf{A} and \textbf{B} first exchange noisy data over the channel $C_p$ (for example, \textbf{A} sends $N_0$ bits to \textbf{B}, and \textbf{B} receives a noisy version of those bits). Then, by public discussion over $C_0$, \textbf{A} and \textbf{B} select $N_1 < N_0$ bits out of the $N_0$ bits that were first exchanged, in such a way that the average error between the $N_1$-long bit string owned by \textbf{A} and the one owned by \textbf{B} is strictly less than $p$.

Advantage Distillation is designed in such a way that the error probability of the channel from \textbf{A} to \textbf{B} decreases more quickly than the error probability of the channel from \textbf{A} to \textbf{E} (and from \textbf{B} to \textbf{E}). A notorious example of Advantage Distillation protocol is the Bit Pair Iteration protocol; \textbf{A} and \textbf{B} send over $C_0$ the parity of each pair of bits of the data they own. When the parity is the same, they retain the first bit; in the other case, they discard the whole pair.

The distillation is made several times until the information sent is likely to have been sent from \textbf{A} to \textbf{B} through a BSC channel $C_\epsilon$ with $\epsilon$ small enough, and the information that \textbf{E} gets was sent through a channel $C_\lambda$ with $\epsilon < \lambda$. After $k$ iterations, \textbf{A} and \textbf{B} share $N_k$ bits with error probability $\epsilon$.

\subsection{Information Reconciliation}

After the step of Advantage Distillation, the bit strings that \textbf{A} and \textbf{B} own still differ. Information Reconciliation aims at correcting these errors by public discussion over $C_0$. \cite{ChaFum06} shows how to modify the Information Reconciliation protocol Cascade \cite{BraSal94} to reduce its hardware implementation to fit into resource-constrained environment. In a nutshell, the Cascade protocol requires \textbf{A} and \textbf{B} to send the parity of blocs of data of increasing size, in such a way that they can correct the few errors remaining with high probability.

\subsection{Privacy Amplification}

\textbf{A} and \textbf{B} now agree on a bit string $S$ of length $N_k$ with very high probability. The aim of Privacy Amplification is to derive a shorter key out of the shared data, on which Eve has no information. For that purpose, \textbf{A} and \textbf{B} agree on a universal hash function from a predefined family of functions, and compute the hash of the bit string. This gives a shorter key $K$ which is the result of the Key Agreement protocol; \cite{CacMau95} proves that \textbf{E} finally does not get any information on $K$.

For practical purposes, the Universal Hash Functions defined in \cite{KaYuSu05} are suited for low hardware requirements.

\subsection{Summary}

These three steps are well known, and enable Key Agreement over a noisy and public channel. However, such a construction is only valid for a passive adversary, \textit{i.e.} when Eve just listens to messages that were sent over the air. In the era of wireless communication, anyone can temper with the data that was sent over a wireless channel, which is the base of packet injection attacks.

The rest of the paper describes our contribution: how to adapt this scheme so that the key establishment protocol described above is resistant to active attacks? 

\section{Integrity $(I)$-codes} \label{sec:iCodes}

In a wireless environment, there is no existing mechanism that prevents an adversary to jam all communication between two devices. Indeed, a powerful white noise can make a Signal-to-Noise Ratio as low as possible. Thus, our goal is not to ensure that no one jams the communication, but to prevent an active adversary to obtain a significant advantage against one of the devices. The sole detection of an attack is thus enough in our model.

We therefore describe a protection system made to detect all intrusion attempts in the communications between \textbf{A} and \textbf{B}, called Integrity Code. These were introduced in \cite{CCRTHS08,CHCRTS06}, and make use of physical means to protect the communication.

Integrity $(I)$-code bits are transmitted in such a way that an adversary can hardly change a bit ``1'' into a ``0''.  Moreover, information is  coded in order to detect the remaining possible bit flipping: from ``0'' to ``1''. Putting these 2 protections together, an adversary cannot modify a message without having a high probability of being detected.

\begin{remark}
Our use of integrity $(I)$-codes enables us to fulfil the non-Simulatability Condition introduced in \cite{DBLP:journals/tit/MaurerW03a}.
\end{remark}

\subsection{Physical Transmission}

The bits are transmitted using the \textbf{On-off keying} technique (OOK). Signal is divided in time-periods of length T. Each bit ``1'' is transmitted as a non-null signal of duration T. Each bit ``0'' corresponds to the absence of signal  during the same amount of time T. 

As the elimination of a non-null electromagnetic signal is very costly, this satisfies the first constraint: preventing the flipping from a ``1'' to a ``0''. 

\begin{assumption}
\label{ass:ook}
It is impossible for an adversary to alter the transmission of a binary ``1'' using OOK.
\end{assumption}

\subsection{Unidirectional Coding}

In order to detect the flipping from a ``0'' to a ``1'', information is coded using a \textbf{Unidirectional Error-Detecting Code} \cite{Berger61}:
\begin{definition} \label{def}
A Unidirectional Error-Detecting  Code is~a triple $(S, C, \alpha)$, satisfying the following conditions:
\begin{enumerate}
\item $S$ is a finite set of possible source states,
\item $C$ is a finite set of binary codewords,
\item $\alpha$ is a source encoding rule $\alpha : S \rightarrow C$, such that:
\begin{itemize}
\item $\alpha$ is an injective function,
\item $C$ respects the ``non-inclusive supports'' property, \textit{i.e.} it is not possible to convert codeword $c \in C$ to another codeword $c' \in C$, such that $c' \neq c$,
without switching at least one bit 1 of $c$ to bit 0.
\end{itemize}
\end{enumerate}
\end{definition}

The ``non-inclusive supports'' property can be restated this way: if $c\in C$ is a binary codeword of length $n$, and $\mathsf{supp}(c) = \left \{i\in \{1, \ldots n\} |  c_i = 1 \right \}$ is the support of $c$, then $\forall c,c'\in C$, the supports of $c$ and $c'$ are not included one into the other, \textit{i.e.} $\mathsf{supp}(c) \not\subset \mathsf{supp}(c')$ and $\mathsf{supp}(c') \not\subset \mathsf{supp}(c)$.

The Manchester coding which encodes bit ``1'' into 10 and bit ``0'' into 01 is a very simple example of unidirectional error-detecting  code. When combined with On-Off Keying, its error-detection rule simply consists in verifying that a codeword contains an equal number of symbols ``0'' and ``1''.

More generally, any binary immutable WOM-code (codes dedicated to Write-Once Memory) permits unidirectional coding. A Write-Once Memory is an array of bits such that once a bit was set to ``1'' it can never be unset again; immutable WOM-codes prevent the rewriting of a message on a Write-Once Memory. To improve the Manchester code, which has a rate of $\frac{1}{2}$, and following \cite{GodCoh86}, we suggest the use of the Berger code. To encode a word $x$ of length $l$, we add $\left\lceil \log l\right\rceil$ bits of redundancy in the following way: the binary weight $w(x)=\sum_{i=1}^{l} x_i$ is computed, and represented in its binary version $w_1, \ldots, w_{\left\lceil \log l\right\rceil}$. The coded version of $x$ is the concatenation of $x$ with $\overline{w_1, \ldots, w_{\left\lceil \log l\right\rceil}}$, \textit{i.e.} $\left (x_1, \ldots, x_l, \overline{w_1}, \ldots, \overline{w_{\left\lceil \log l\right\rceil}}\right )$ \footnote{The notation $\overline{a}$ is the binary negation of $a$.}. The Berger code works because if $\mathsf{supp}(x) \subset \mathsf{supp}(x')$, then $w(x) \leq w(x')$, and $\mathsf{supp}\left (\overline{w_1}, \ldots, \overline{w_{\left\lceil \log l\right\rceil}} \right) \not\subset \mathsf{supp}\left ( \overline{w'_1}, \ldots, \overline{w'_{\left\lceil \log l\right\rceil}} \right )$.

\begin{remark}
The idea of unidirectional coding was introduced by \cite{DBLP:conf/eurocrypt/Maurer97} in the same context.
\end{remark}

\section{Key Agreement Through Presence} \label{sec:ka-pres}

\subsection{The Model}

Here is the description of the model for which we design the protocol. It is based on the facts described previously: communication between wireless devices is public, any adversary can make the communication unreadable, it is not possible to make expensive computation with cheap devices. Therefore, the following hypothesis are made:
\begin{itemize}
	\item \textbf{A} is a low-cost device with limited computation and memory possibilities;
	\item \textbf{B} is a wireless sensor \textit{i.e.} a communicating device that has reasonable computing hardware;
	\item The two devices \textbf{A} and \textbf{B} are \emph{in presence}, which means that they are communicating with each other, and not with a third party \textbf{E};
	\item \textbf{E} can hear everything that \textbf{A} and \textbf{B} send;
	\item \textbf{E} is able to emit at the same time an electromagnetic signal.
\end{itemize}

This last item is the main difference between the existing protocols and the following: we here consider \emph{active adversaries}.

\begin{definition} \label{def:resistant}
Let $C$ be a channel between \textbf{A} and \textbf{B}, and \textbf{E} be an adversary such that:
\begin{itemize}
	\item Transmission of a message $s= (x_1, \ldots, x_n)\in \{0,1\}^n$ from \textbf{A} to \textbf{B} without interference of \textbf{E} is noiseless.
	\item Transmission of $s$ from \textbf{A} to \textbf{B} with intervention of \textbf{E} leads to the reception of $\Phi_E(s)=s'=(x'_1, \ldots, x'_n)$.
	\item A failed transmission leads to a state $\bot$ for \textbf{A} and \textbf{B}.
\end{itemize}
$C$ is $\epsilon$-resistant against an active adversary if except with probability less than $\epsilon$, $\forall s \in \{0,1\}^n, s=\Phi_E(s)$ or \textbf{A} and \textbf{B} are in the state $\bot$.
\end{definition}

Such a channel is such that, after a transmission, either \textbf{A} and \textbf{B} possess the same message $s$, or \textbf{A} and \textbf{B} know that the transmission was a failure.

\subsection{Rewriting the Three Steps}

As we mentioned it in Section \ref{sec:formal}, there are two channels for \textbf{A} and \textbf{B} to communicate. The first one is $C_p$, the second $C_0$.

\begin{enumerate}
\item The messages that are sent over the channel $C_0$ are error-less thanks to error correction techniques. To eliminate an active adversary's chances of tempering with this channel, we add a fourth step called \highlight{Integrity Verification} after the three enumerated in Section~\ref{sec:keyAgr}, described hereafter.
\item In the classical key agreement protocol, the channel $C_p$ between \textbf{A} and \textbf{B} (resp. \textbf{A} and \textbf{E}) is usually modeled as a BSC channel with error probability $p_{AB}$ (resp. $p_{AE}$). If the adversary is active during the first phase, then the effect is an increase of $p_{AB}$ without a change on $p_{AE}$. However, the Advantage Distillation step finally leads to a new error probability $p'_{AB}$ that is lower than $p_{AE}$ independently of the initial situation. Therefore, thanks to the final Integrity Verification, an active adversary cannot gain an advantage at this step.
\end{enumerate}

\subsection{Validating the Agreement}

The final verification step permits to ensure that the key agreement protocol was not perturbed by an active adversary. For that, the idea is to check that all the messages sent and received by \textbf{A} and \textbf{B} were the same, using a protection technique on the verification message.

Note $\mathcal{M}$ the set of all the messages that were emitted by both devices, in their order of apparition. We expect \textbf{B} to continuously save $\mathcal{M}$. At the end of the protocol, \textbf{A} will send to the wireless sensor \textbf{B} the identifier of a function $h$ taken from a family of hash functions, together with $\alpha \left ( h (\mathcal{M}) \right )$ where $\alpha$ is the source encoding rule defined in Definition~\ref{def}.

To reduce memory usage, \textbf{A} can compute $h(\mathcal{M})$ in an incremental way, by $x_{n+1}=h(x_n || m_{n+1})$ with $m_i$ the $i$-th message transmitted over the channel, and $x_i$ the hash of the $i$ first elements. $||$ is the concatenation operator. 

\begin{figure}[htb]
  \centering
	\includegraphics[width=0.9\columnwidth ]{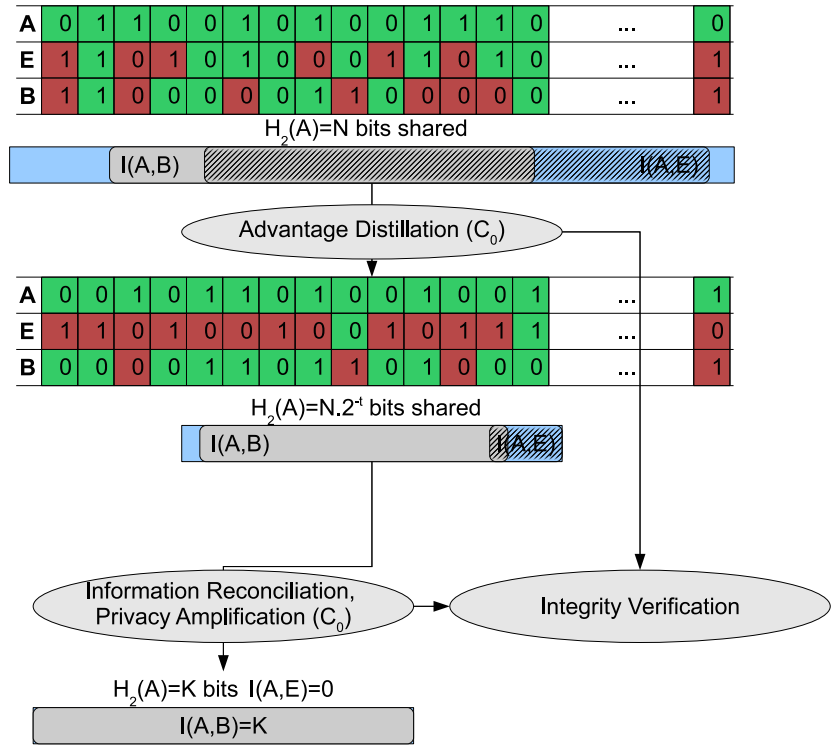} 
\caption{The global scheme, illustrated}
\label{fig:glob}
\end{figure}

We therefore suggest the following order for the global scheme, which is illustrated in Fig.~\ref{fig:glob}.
\begin{enumerate}
	\item \textbf{A} chooses the hash function $h$ from a family of hash functions;
	\item \textbf{B} sends to \textbf{A} a bit stream using $C_{p_{AB}}$;
	\item \textbf{A} and \textbf{B} proceed to Advantage Distillation, Information Reconciliation, and Privacy Amplification;
	\item \textbf{A} sends to \textbf{B} the identifier of $h$;
	\item \textbf{A} and \textbf{B} do the Integrity Verification step: \textbf{B} sends to \textbf{A} the message $\alpha \left ( h(\mathcal{M})\right )$ where $\mathcal{M}$ designates all the messages that were sent over $C_0$, using On-Off Keying (over $C_0$). 
\end{enumerate}

\subsection{The Noiseless Shielded Channel}

We here deliver the statement made in Section \ref{sec:formal}: with simple tools, to achieve a channel that is noiseless and integrity resistant against the intrusion of an active adversary.

The channel designed so far complies with Definition~\ref{def:resistant}, as this is expressed in the following formalization:
	Let \textbf{A} and \textbf{B} be a sender and a receiver; let $n,t_1,t_2\in \mathbb{N}$ with $n\geq t_1$ and $t_2 \geq t_1$, $h:\{0,1\}^n \rightarrow \{0,1\}^{t_1}$ be a hash function, and $\alpha: \{0,1\}^{t_1} \rightarrow \{0,1\}^{t_2}$ a source encoding rule following Definition \ref{def}.
	
	\textbf{A} emits a message $s\in\{0,1\}^n$ to \textbf{B} using On-Off Keying, by sending $S = s || \alpha(h(s))$. At the reception of $S' = s_1' || s_2'$ with $|s_1'|=n$, \textbf{B} checks that $s_2' = \alpha(h(s_1'))$. If this test fails, then \textbf{B} emits a standard message expressing failure. If not, \textbf{B} uses the now shared key to validate the agreement.

\begin{prop}	
	The scheme described in the previous paragraph gives a channel $C$ that is $\epsilon$-resistant against an active adversary, where $$\epsilon=\Pr_{x,x'}\left [h(x)=h(x')\right ]$$ is the collision probability of $h$.
\end{prop}

\demo
Two cases need to be considered : either \textbf{E} does not intervene, or \textbf{E} tries to alter the communications.
In the first case, we obviously have $S=S'$, which also gives $s=s_1'$ which was the desired result.

In the second case, note that, thanks to OOK (see Assumption~\ref{ass:ook}), the only action \textbf{E} can do is to change a ``0'' that was sent into a ``1''.
\begin{itemize}
\item If \textbf{E} alters $\alpha(h(s))$ into $s_2'$, using the unidirectional property of $\alpha$, the equality $s_2' = \alpha(h(s_1'))$ is never achieved.
\item If \textbf{E} alters $s$ into $s'$, but not $\alpha(h(s))$ then \textbf{E} wins only if $h(s)=h(s')$, \textit{i.e.} with probability less than $\epsilon$.
\end{itemize}
This shows that the alteration of a message by $\textbf{E}$ is detected with probability greater than $1-\epsilon$. Therefore the channel is $\epsilon$-resistant against an active adversary.
\findemo

In our application, an active \textbf{E} can alter the agreement on the hash function $h$. If this happens, then \textbf{A} owns a function $h_A$ and \textbf{B}, $h_B$. With this kind of advantage, \textbf{E} must nonetheless change $\mathcal{M}_A, \mathcal{M}_B$ into $\mathcal{M}'_A, \mathcal{M}'_B$, with the properties $h_A(\mathcal{M}_A) = h_B(\mathcal{M}'_A)$ and $h_A(\mathcal{M}'_B) = h_B(\mathcal{M}_B)$.
Moreover, to successfully interfere in the communication, an active \textbf{E} must change ``on the fly'' messages that are sent by \textbf{A} and \textbf{B} such that the final hashes collide, with no knowledge of the future messages to be sent, and with the constraint $\mathsf{supp}(x) \subset \mathsf{supp}(x')$, \textit{i.e.} \textbf{E} can only change ``0'' into ``1''. This makes her task even harder.

\begin{remark}
Our new approach does not resist to an active adversary issuing a low-energy DoS attack to invalidate all key exchanges. As mentioned earlier, our goal is not to prevent DoS attacks. 
\end{remark}

\section{Conclusion} \label{sec:concl}

This paper describes a method to establish a key with a low cost wireless device. Starting from the classical key agreement methods, we provide the tools to achieve the integrity mechanisms necessary in order to cope with active adversaries. Using integrity $(I)$-codes - a modulation method that prevents to switch from a ``1'' to a ``0'', combined with unidirectional coding, we add a fourth step that detects intrusion in the communication.

This paper finally focuses on the computation cost so that devices with very few logical gates can instantiate this protocol. Indeed, the device needs only to implement a few functions for the protocol to work: 
\begin{itemize}
\item A parity evaluator -- for the Advantage Distillation and Information Reconciliation steps,
\item A universal hash function, for Privacy Amplification,
\item A unidirectional coding scheme, for Integrity Verification,
\item A binary comparator.
\end{itemize}
The universal hash function is here the most gate-consuming element, and can be designed in roughly 640 gates following \cite{Yuksel04}. The universal coding scheme, that uses a Berger code, only requires to compute a binary weight, and a logical negation. For key length of about $64$ bits, this can be done in about 320 gates. Finally, the overall complexity of such a device is of the order of 1000 logical gates.

This makes way for the production of large amounts of low-cost tags allowing secure communication. 

~\\

\noindent {\bf Acknowledgments}
The authors thank the referees for their comments.

\bibliographystyle{IEEEbib}
\bibliography{communications2}

\begin{thebibliography}{10}

\bibitem{ChaFum06}
H.~Chabanne and G.~Fumaroli,
\newblock ``Noisy cryptographic protocols for low-cost rfid tags,''
\newblock {\em IEEE Transactions on Information Theory}, vol. 52, no. 8, pp.
  3562--3566, Aug. 2006.

\bibitem{GanMau94}
M.J. Gander and U.M. Maurer,
\newblock ``On the secret-key rate of binary random variables,''
\newblock in {\em IEEE International Symposium on Information Theory, 1994.}

\bibitem{BraSal94}
G.~Brassard and L.~Salvail,
\newblock ``Secret-key reconciliation by public discussion,''
\newblock in {\em EUROCRYPT '93}. Springer-Verlag.

\bibitem{CacMau95}
C.~Cachin and U.~M. Maurer,
\newblock ``Linking information reconciliation and privacy amplification,''
\newblock in {\em {EUROCRYPT'94}}. Springer Berlin.

\bibitem{KaYuSu05}
J.-P. Kaps, K.~Yuksel, and B.~Sunar,
\newblock ``Energy scalable universal hashing,''
\newblock {\em IEEE Transactions on Computers}, vol. 54, no. 12, pp.
  1484--1495, Dec. 2005.

\bibitem{CCRTHS08}
S.~Capkun, M.~Cagalj, R.~Rengaswamy, I.~Tsigkogiannis, J.-P. Hubaux, and
  M.~Srivastava,
\newblock ``Integrity codes: Message integrity protection and authentication
  over insecure channels,''
\newblock {\em Dependable and Secure Computing, IEEE Transactions on}, vol. 5,
  no. 4, pp. 208--223, Oct.-Dec. 2008.

\bibitem{CHCRTS06}
M.~Cagalj, J.-P. Hubaux, S.~Capkun, R.~Rengaswamy, I.~Tsigkogiannis, and
  M.~Srivastava,
\newblock ``Integrity (i) codes: Message integrity protection and
  authentication over insecure channels,''
\newblock {\em IEEE Symposium on Security and Privacy}, pp. 280--294, 2006.

\bibitem{DBLP:journals/tit/MaurerW03a}
Ueli~M. Maurer and Stefan Wolf,
\newblock ``Secret-key agreement over unauthenticated public channels ii: the
  simulatability condition,''
\newblock {\em IEEE Transactions on Information Theory}, vol. 49, no. 4, pp.
  832--838, 2003.

\bibitem{Berger61}
J.M. Berger,
\newblock ``A note on error detection codes for asymmetric channels,''
\newblock {\em Information and Control}, vol. 4, no. 1, pp. 68 -- 73, 1961.

\bibitem{GodCoh86}
P.~Godlewski and G.~D. Cohen,
\newblock ``Some cryptographic aspects of womcodes,''
\newblock in {\em CRYPTO'85}. Springer-Verlag.

\bibitem{DBLP:conf/eurocrypt/Maurer97}
Ueli~M. Maurer,
\newblock ``Information-theoretically secure secret-key agreement by not
  authenticated public discussion,''
\newblock in {\em {EUROCRYPT'97}}, 1997.

\bibitem{Yuksel04}
K.~Yuksel,
\newblock ``Universal hashing for ultra-low-power cryptographic hardware
  applications,''
\newblock M.S. thesis, Worcester Polytechnic Institute, 2004.

\end{thebibliography}

\end{document}